\documentclass[12pt, 
               pra, 
               preprintnumbers, 
               superscriptaddress,
               showpacs,
               a4paper]{revtex4}

\usepackage{amsfonts}
\usepackage{amsmath}
\usepackage{bm}
\usepackage[dvips]{graphicx}
\usepackage{color}
\usepackage{hyperref}
\usepackage{epsfig}

\DeclareMathAlphabet{\mathpzc}{OT1}{pzc}{m}{it}

\newcommand{\la}{\langle}
\newcommand{\ra}{\rangle}

\begin{document}

\author{Arina Sytcheva}
\email[e-mail: ]{arina.sytcheva@cfel.de}
\affiliation{Center for Free-Electron Laser Science, DESY, 22607 Hamburg, Germany}
\author{Stefan Pabst}
\affiliation{Center for Free-Electron Laser Science, DESY, 22607 Hamburg, Germany}
\affiliation{Department of Physics, University of Hamburg, 20355 Hamburg, Germany}
\author{Sang-Kil Son}
\affiliation{Center for Free-Electron Laser Science, DESY, 22607 Hamburg, Germany}
\author{Robin Santra}
\email[e-mail: ]{robin.santra@cfel.de}
\affiliation{Center for Free-Electron Laser Science, DESY, 22607 Hamburg, Germany}
\affiliation{Department of Physics, University of Hamburg, 20355 Hamburg, Germany}

\title{Enhanced nonlinear response of Ne$^{8+}$ to intense ultrafast x rays}

\date{\today}

%
%

\begin{abstract}

We investigate the possible reasons for the discrepancy between the theoretical two-photon ionization cross section, $\sim 10^{-56}$~$\text{cm}^4 \text{s}$, of Ne$^{8+}$ obtained within the perturbative nonrelativistic framework for monochromatic light [J. Phys. B {\bf 34}, 4857 (2001)] and the experimental value, $7 \times 10^{-54}$ $\text{cm}^4 \text{s}$, reported in [Phys. Rev. Lett. {\bf 106}, 083002 (2011)] at a photon energy of 1110 eV. To this end, we consider Ne$^{8+}$ exposed to deterministic and chaotic ensembles of intense x-ray pulses. The time-dependent configuration-interaction singles (TDCIS) method is used to quantitatively describe nonlinear ionization of Ne$^{8+}$ induced by coherent intense ultrashort x-ray laser pulses. The impact of the bandwidth of a chaotic ensemble of x-ray pulses on the effective two-photon ionization cross section is studied within the lowest nonvanishing order of perturbation theory. We find that, at a bandwidth of 11 eV, the effective two-photon ionization cross section of Ne$^{8+}$ at a photon energy of 1110 eV amounts to $5 \times 10^{-57}$ and $1.6 \times 10^{-55}$ $\text{cm}^4 \text{s}$ for a deterministic ensemble and a chaotic ensemble, respectively. We show that the enhancement obtained for a chaotic ensemble of pulses originates from the presence of the one-photon 1$s^2$--1$s$4$p$ resonance located at 1127 eV. Using the TDCIS approach, we also show that, for currently available radiation intensities, two-photon ionization of a 1$s$ electron in neutral neon remains less probable than one-photon ionization of a valence electron.

\end{abstract}

\maketitle


\section{Introduction} \label{sec:introduction}
Modern highly intense x-ray free-electron lasers (XFELs), such as the Linac Coherent Light Source (LCLS) at SLAC National Accelerator Laboratory, Stanford, USA \cite{emma}, and the SPring-8 {\AA}ngstr\"om Compact Free-Electron Laser (SACLA), Harima, Japan \cite{pile2011x}, deliver both soft and hard x-ray radiation. FLASH at DESY, Hamburg, Germany \cite{ackermann2007operation}, operates in the VUV and soft x-ray regimes, and the European XFEL \cite{europeXFEL}, which is under construction, is planned to deliver photon energies up to 12 keV. These facilities offer possibilities to explore inner-shell electron dynamics and nonlinear response of atoms and molecules to intense x-ray radiation (see for example \cite{young-nature, doumy-prl2011, richter2009extreme, rudenko2008recoil, moshammer2007few, liu2008strong}).

The present theoretical work is triggered by a recent experiment on nonlinear ionization of neon atoms performed at the LCLS \cite{doumy-prl2011}.
The experiment utilized the capability of the LCLS to produce unprecedentedly intense x-ray beams, with up to $\sim$\nolinebreak10$^{12}$ x-ray photons in a $\sim$\nolinebreak100 fs pulse with a peak intensity of $\sim 10^{17}$ \nolinebreak $\text{W/cm}^2$. Within a single pulse the initially neutral target absorbed multiple photons yielding a variety of ion species in different electronic configurations. At a photon energy of 1110 eV, which is below the $K$-shell threshold of Ne$^{8+}$, Doumy {\it et al.} \cite{doumy-prl2011} observed production of hydrogen-like neon, Ne$^{9+}$. 
The Ne$^{9+}$/Ne$^{8+}$ ratio was observed to depend quadratically on the peak intensity, which is consistent with nonlinear two-photon ionization of Ne$^{8+}$. Nevertheless, the two-photon ionization cross section, deduced from this experimental observation with the help of a rate-equation model, is $7 \times 10^{-54}$ \nolinebreak $\text{cm}^4 \text{s}$, which is two orders of magnitude higher than the value obtained within perturbation theory \cite{nov-hasp2001, koval}.

In the present paper, we focus on the following points: (i) the discrepancy between the observed \cite{doumy-prl2011} and theoretically predicted \cite{nov-hasp2001} two-photon ionization cross section values of Ne$^{8+}$; (ii) the possibility of two-photon ionization of a 1$s$ electron in neutral neon below the $K$-shell threshold of neon. 

To describe the nonlinear interaction of Ne$^{8+}$ and neutral neon with an intense coherent ultrashort x-ray pulse, we adopt the time-dependent configuration-interaction singles (TDCIS) method---a nonperturbative {\it ab initio} multichannel approach \cite{rohr-pra74, pabst, pabst-prl, saalfrank}. TDCIS allows for pulses of arbitrary shape and peak intensity, and provides an intuitive picture of the electron dynamics induced by a light pulse of finite duration. Correlation effects between the ejected photoelectron and the remaining ion are included via exact treatment of the Coulomb interaction \cite{pabst-prl}. Going beyond the standard single-active electron approximation \cite{rohr-pra74, sae-kulander1991}, the TDCIS model accounts for the coupling between different excitation (ionization) channels.

In our study, we employ the TDCIS method implemented in the \textsc{xcid} code \cite{xcid}. To eliminate spurious reflections, which appear when the electronic wave packet reaches the boundary of the numerical grid, we apply absorbing boundaries through the inclusion of a complex absorbing potential (CAP) \cite{cap-riss, cap-santra}. Implemented within the framework of TDCIS, the CAP provides a measure for the ionization probability for the outgoing electron. The ionization probability, given by the diagonal components of the reduced ion density matrix (IDM), is used in this work for calculating the generalized two-photon ionization cross section.

For nonlinear light-matter interaction the spectral and temporal shape of the pulse is a crucial factor \cite{loudon, saldin2000physics}. The rate of simultaneous absorption of two photons depends on the statistics of the exciting field \cite{lambropoul-PhysRev1968, lambropoul-PhysRev1966, mollow-PhysRev1968}. Present XFELs have a coherence time that is much shorter than the pulse duration and can be considered as chaotic \cite{Dattoli-1984, saldin2010statistical}. For a chaotic ensemble of pulses \cite{mollow-PhysRev1968} with a finite bandwidth and a short coherence time \cite{ho2008diffraction},  within the lowest nonvanishing order of perturbation theory (LOPT), the effective two-photon ionization cross section can be written as a convolution of the monochromatic two-photon cross section and the spectral distribution function. We investigate the effect of finite coherence time on the two-photon ionization cross section by using a Gaussian spectral distribution function. The monochromatic two-photon ionization cross section we calculate within the Hartree-Fock-Slater (HFS) model \cite{slater}, implemented within the \textsc{xatom} code \cite{son-xatom2011, xatom}.

The results to be presented here indicate that the treatment of XFEL radiation as a chaotic finite-bandwidth ensemble of pulses, rather than a deterministic ensemble of pulses, is likely to be capable of explaining the enhanced two-photon ionization cross section reported in Ref. \cite{doumy-prl2011}.

The paper is organized as follows: we outline the theoretical approaches in Sec.~\ref{sec:theory}, present details on the numerical implementation and the obtained results in Sec.~\ref{sec:results} and draw conclusions in Sec.~\ref{sec:conclusions}. Atomic units are used throughout, unless otherwise noted.

\section{Theory}\label{sec:theory}

\subsection{Two-photon ionization cross section for a coherent finite pulse} \label{sec:tdcis}
A detailed description of our implementation of the TDCIS method can be found in Ref.~\cite{pabst}. Briefly, we construct the electronic wave packet in an atom as a linear combination of the Hartree-Fock ground state $| \Phi_0 \rangle$ and one-particle--one-hole (1p--1h) excitations $|\Phi_i^a \rangle$, 
\begin{equation}
| \Psi,t \rangle = \alpha_0(t) | \Phi_0 \rangle + \sum_i \sum_a \alpha_i^a (t) | \Phi_i^a \rangle,
\label{eq:wavepacket}
\end{equation}
where
\begin{equation}
| \Phi_i^a \rangle = \frac{1}{\sqrt2} \{ \hat{c}_{a \uparrow}^{\dagger} \hat{c}_{i \uparrow}^{\phantom\dagger} + \hat{c}_{a \downarrow}^{\dagger} \hat{c}_{i \downarrow}^{\phantom\dagger} \} | \Phi_0 \rangle.
\end{equation}
Here, $i,\ j, \ldots$ label orbitals occupied  in $| \Phi_0 \rangle$, whereas unoccupied (virtual) orbitals are marked by $a,\ b, \ldots$ The operators $\hat{c}_{p \sigma}^{\dagger}$ and $\hat{c}_{p \sigma}$ create and annihilate, respectively, electrons in a spin orbital of the modified Fock operator $\hat{F}_{\text{CAP}}=\hat{F}-i \eta \hat{W}$, which consists of the Fock operator $\hat{F}$ and the CAP in the form $-i \eta \hat{W}$. The spin states are designated with $\sigma$. In the electric dipole approximation, the nonrelativistic Hamiltonian of the atom interacting with the x-ray field is given by
\begin{equation}
\hat{H}=\hat{F}_{\text{CAP}} + \hat{V}_C - \hat{V}_{\text{HF}} - E_{\text{HF}} - {\cal E} (t) \hat{z},
\label{eq:hamil} 
\end{equation} 
where $\hat{V}_C$ stands for the electron-electron Coulomb interaction, $\hat{V}_{\text{HF}}$ and $E_{\text{HF}}$ are  the Hartree-Fock mean-field potential and ground-state energy, respectively, $\hat{z}$ is the dipole operator, and ${\cal E} (t)$ is the electric field of the intense ultrashort laser pulse, which is assumed to be linearly polarized along the $z$ axis. By substituting the wave function given by Eq. (\ref{eq:wavepacket}) into the time-dependent Schr\"odinger equation, one gets a set of coupled ordinary differential equations for the coefficients $\alpha_0(t)$ and $\alpha_i^a(t)$. 

Using the state $|\Psi, t \rangle$, we construct the reduced density matrix of the residual ion produced in the photoionization process, 
\begin{equation}
\hat{\rho}(t) = \text{Tr}_a \left[\,| \Psi, t \rangle \langle \Psi, t |\,\right],
\label{eq:idm}
\end{equation}
\begin{equation}
\rho_{ij}(t) = \sum_{a,b} \alpha_i^a (t) [\alpha_j^b(t)]^{*} o_{ab},
\end{equation}
where $o_{ab}$ stands for the overlap between eigenfunctions of $\hat{F}_{\text{CAP}}$. The CAP is only active at large distances from the atom, and, hence, affects only virtual orbitals. Application of the CAP is equivalent to attenuation of the wave packet when it reaches the boundary of the numerical grid \cite{kosloff}. Because of the CAP, the norm of the wave packet from Eq. (\ref{eq:wavepacket}) as well as the norm of the reduced ion density matrix (\ref{eq:idm}), are not conserved and decrease as ionization proceeds. In order to compensate for this loss of norm in the IDM, one has to introduce a correction \cite{pabst, rohr-pra79}:
\begin{equation}
\delta \rho_{ij}(t) = 2 \eta e^{i(\varepsilon_i-\varepsilon_j)t } \int\limits_{-\infty}^{~t} dt' \sum_{a,b} w_{ba} \alpha_i^a (t') [\alpha_j^b (t')]^* e^{-i(\varepsilon_i-\varepsilon_j)t' },
\label{eq:correction1}
\end{equation}
with the $\varepsilon_i$ being the orbital energies and $w_{ba}$ the matrix elements of the CAP operator $\hat{W}$. In the limit $t \rightarrow \infty$, {\it i.e.}, after the ionizing pulse is over, a diagonal component of the corrected IDM, $\rho_{i} + \delta \rho_{i} ( \equiv \rho_{ii} + \delta \rho_{ii})$, can be thought of as the excitation probability from an occupied orbital $i$. Under the conditions considered here, the uncorrected $\rho_{i}$ vanishes for sufficiently long time after the pulse is over, indicating that the photoelectron is completely absorbed by the CAP. Conversely, the IDM correction, $\delta \rho_{i}$, approaches a constant value at $t \rightarrow \infty$ and can be interpreted as the ionization probability of an electron from orbital $i$.

The ionization probability per unit time due to direct absorption of $N$ photons (in $\text{s}^{-1}$) is given by $\sigma^{(N)} J ^N$ where $J$ is the photon flux in number of photons per cm$^2$ per second. This allows for a definition of an effective two-photon ionization cross section for a coherent pulse centered at a mean photon energy $\omega_{\text{in}}$ with a bandwidth of $\Delta \omega_{\text{p}}$  
\begin{equation}
\sigma^{(2)}_{\rm coh} (\omega_{\text{in}}, \Delta \omega_{\text{p}})=\frac{ \lim \limits_{t \rightarrow \infty} \delta \rho_{i} (t)}{\int\limits_{-\infty}^{\infty} dt \ J(t)^2}.
\label{eq:cohcs} 
\end{equation}
The quantities $\omega_{\text{in}}$  and $\Delta \omega_{\text{p}}$ enter the right-hand side of Eq.~(\ref{eq:cohcs}) implicitly through the IDM correction $ \delta  \rho_{i}$, obtained using the Hamiltonian from Eq.~(\ref{eq:hamil}), and the flux $J(t)$. The definition of Eq.~(\ref{eq:cohcs}) is valid provided the ground state is not depleted, {\it i.e.}, in the perturbative limit.

\subsection{Two-photon ionization cross section for chaotic fields}

When defining the cross section in Eq. (\ref{eq:cohcs}) we assume that the x-ray pulse is well defined (deterministic). In general, the radiation produced by an XFEL operating in the self-amplified spontaneous emission (SASE) regime is chaotic with respect to fluctuations in the electric field. The simplest way to account for the XFEL chaoticity is to recall that the $N$-photon ionization rate, within the lowest nonvanishing order of perturbation theory (LOPT), is proportional to $N!J^N$, which amounts to effective doubling $(2!)$ of the cross section value for two-photon ionization \cite{loudon}. This factor of two cannot explain  the discrepancy found in Ref. \cite{doumy-prl2011}. The most rigorous and accurate way to simulate the experimental situation would be by introducing an appropriate stochastic model \cite{pulse-reiche, pulse-pfeifer} for the radiation and solving the TDCIS equations many times using an ensemble of realistic pulses. Afterwards one would have to average the results over all members of the ensemble. However, this approach is computationally very costly. 

Here, we follow the result of Mollow \cite{mollow-PhysRev1968} who showed that within the second-order perturbation theory for a field consisting of finite chaotic pulses, the transition rate due to two-photon absorption during the pulse can be expressed in terms of the spectral first-order field correlation function. In case of a finite-bandwidth field and near an intermediate resonance of the target atom the two-photon ionization cross section for an incoherent pulse can be cast in the form: 
\begin{equation}
\sigma^{(2)}_{\rm incoh} (\omega_{\text{in}}, \Delta \omega_{\text{p}})= 2 \int \limits_{-\infty}^{\infty} d \omega \ \sigma_{\text{LOPT}}^{(2)} (\omega) F(\omega, \omega_{\text{in}}, \Delta \omega_{\text{p}}),
\label{eq:incohcs}
\end{equation}
where $F(\omega, \omega_{\text{in}}, \Delta \omega_{\text{p}})$ is the normalized spectral distribution function and  $\sigma_{\text{LOPT}}^{(2)}$ is the result of the LOPT for monochromatic radiation \cite{koval, nov-hasp2001, saenz}:
\begin{equation}
\sigma_{\text{LOPT}}^{(2)} (\omega) = \pi  (4 \pi \alpha \omega)^2 \sum \limits_{f} \delta(\omega_{f} - \omega_{g} - 2 \omega) \bigg| \sum \limits_{l} \frac{\la f | z | l \ra \la l | z | g \ra}{\omega_g + \omega - \omega_l + i \Gamma_l/2} \bigg|^2,
\label{eq:lopt} 
\end{equation}
with $\alpha$ being the fine-structure constant. In Eq.~(\ref{eq:lopt}), $|f \ra$, $|l \ra$ and $|g \ra$ stand for final, intermediate and ground states, respectively. $\Gamma_l$ accounts for the natural linewidth of the intermediate states~$ |l \ra$; $\omega_g$ and $\omega_l$ denote energies of the ground and intermediate states, respectively. Note that the factor of 2 in Eq. (\ref{eq:incohcs}) accounts for the enhancement of two-photon absorption from a single-mode chaotic field \cite{loudon}. 
 
The spectral distribution of a single XFEL pulse is very spiky and random \cite{krinsky2003analysis, saldin2000physics}. Averaged over many shots the spectral distribution can be taken as a normalized Gaussian \cite{vartanyants2011coherence, galayda2010x},
\begin{equation}
F(\omega, \omega_{\text{in}}, \Delta \omega_{\text{p}}) = \frac{2 \sqrt{\ln 2}}{\sqrt{\pi} \Delta \omega_{\text{p}} } \exp \Bigg( - 4 \ln 2 \bigg( \frac{\omega-\omega_{\text{in}}} {\Delta \omega_{\text{p}}}  \bigg)^2 \Bigg).
\label{eq:distribution}
\end{equation}

The result given by Eq.~(\ref{eq:incohcs}) can be understood as a nonlinear atomic response to a spectral range of uncorrelated modes. Here, the atomic response to the individual frequencies is summed incoherently. In contrast, Eq.~(\ref{eq:cohcs}) represents nonlinear atomic response to a coherent pulse. In the next section, we apply Eqs.~(\ref{eq:cohcs}) and (\ref{eq:incohcs}) to calculate effective two-photon ionization cross sections of Ne$^{8+}$ in the photon-energy range below its $K$-edge.  

\section{Results and Discussion}  \label{sec:results}

We start our numerical study with the nonlinear atomic response of Ne$^{8+}$ to a deterministic coherent pulse using TDCIS implemented in the \textsc{xcid} code \cite{xcid}. We obtain converged results by using a nonuniform radial grid extending from $r$=0 to $r$=80~a.u.\ with 1000 grid points and a pseudospectral-grid parameter $\zeta=0.461$ \cite{pabst}. Under these conditions, there is an almost uniform orbital energy spacing of about 0.3 a.u.\ across a wide energy range (up to 150 a.u.) for the final states of the outgoing electron. The CAP starts at $r$=50~a.u. We use a CAP strength $\eta = 0.002$ a.u., which makes the energy levels broad enough to describe the quasicontinuum. In this range of $\eta$, we satisfy the stationarity condition with respect to $\eta$: $\partial [ \lim \limits_{t \rightarrow \infty} \delta \rho_{1s} (t)] / \partial \eta=0$, where $\rho_{1s}$ denotes the diagonal component of the IDM corresponding to the $1s$ orbital. The positions of $1s^2$--$1s \, np$ resonances are obtained with an accuracy of 0.03 a.u. and the one-photon ionization potential of Ne$^{8+}$ equals 43.9 a.u. (1194.1 eV). For the comparison, the experimental value of the ionization potential of Ne$^{8+}$ is 1195 eV \cite{xraydatabooklet}. We account for angular momenta of the outgoing electron up to $l_{\text{max}}=2$. The laser pulse is given by ${\cal E}(t)= {\cal E}_0 \exp \lbrace -2 \ln 2 (t/\tau_{\text{p}})^2 \rbrace \cos (\omega_{\text{in}}t)$, where $\tau_{\text{p}}$ is the full-width-at-half-maximum (FWHM) duration of the pulse intensity \footnote{The pulse duration given in femtoseconds is inversely related to the bandwidth of the pulse given in eV as $\Delta \omega_{\text{p}}=1.8/\tau_{\text{p}}$.}, ${\cal E}_0$ is the peak electric field.

\begin{figure}[ht]
\includegraphics[width=0.75\textwidth]{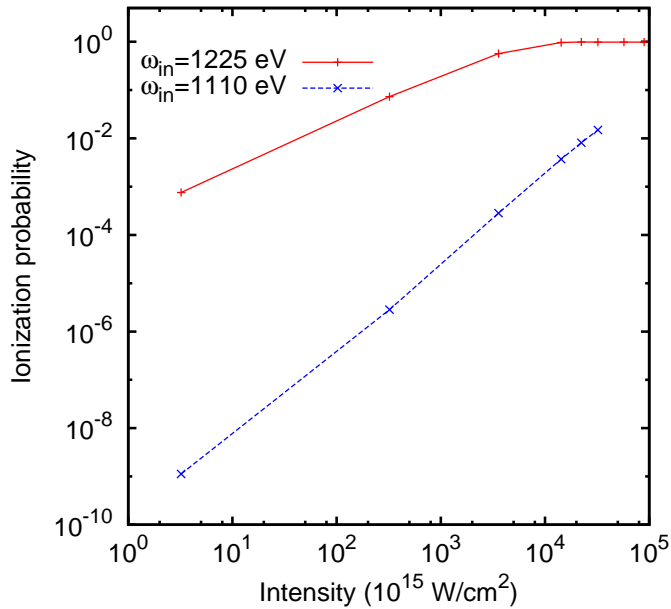}  
\caption{(Color online) Intensity dependence of the ionization probability of Ne$^{8+}$, given by the diagonal IDM correction $\delta \rho_{1s}$, at photon energies of 1225 eV (above the single-photon ionization threshold) and 1110 eV (below the single-photon ionization threshold). A deterministic pulse of 6-eV bandwidth (FWHM) is used.}
\label{fig:respond}
\end{figure}

In Fig.~\ref{fig:respond}, we show how the $1s$ ionization probability depends on intensity at two different photon energies used in the experiment \cite{doumy-prl2011}, below (1110 eV) and above (1225 eV) the one-photon ionization threshold. For the calculation we use a coherent pulse with a FWHM bandwidth of 6 eV. We can see that in double logarithmic scale the slope of the curve corresponding to 1110 eV is 2, while that for 1225 eV the slope below saturation is 1. This reflects the fact that at 1225 eV, $1s$ ionization is a one-photon process, whereas at 1110 eV, it is  a two-photon process. Above $\sim 3\times 10^{18}$ W/cm$^2$, depletion of the ground state becomes substantial.

\begin{figure}[ht]
\includegraphics{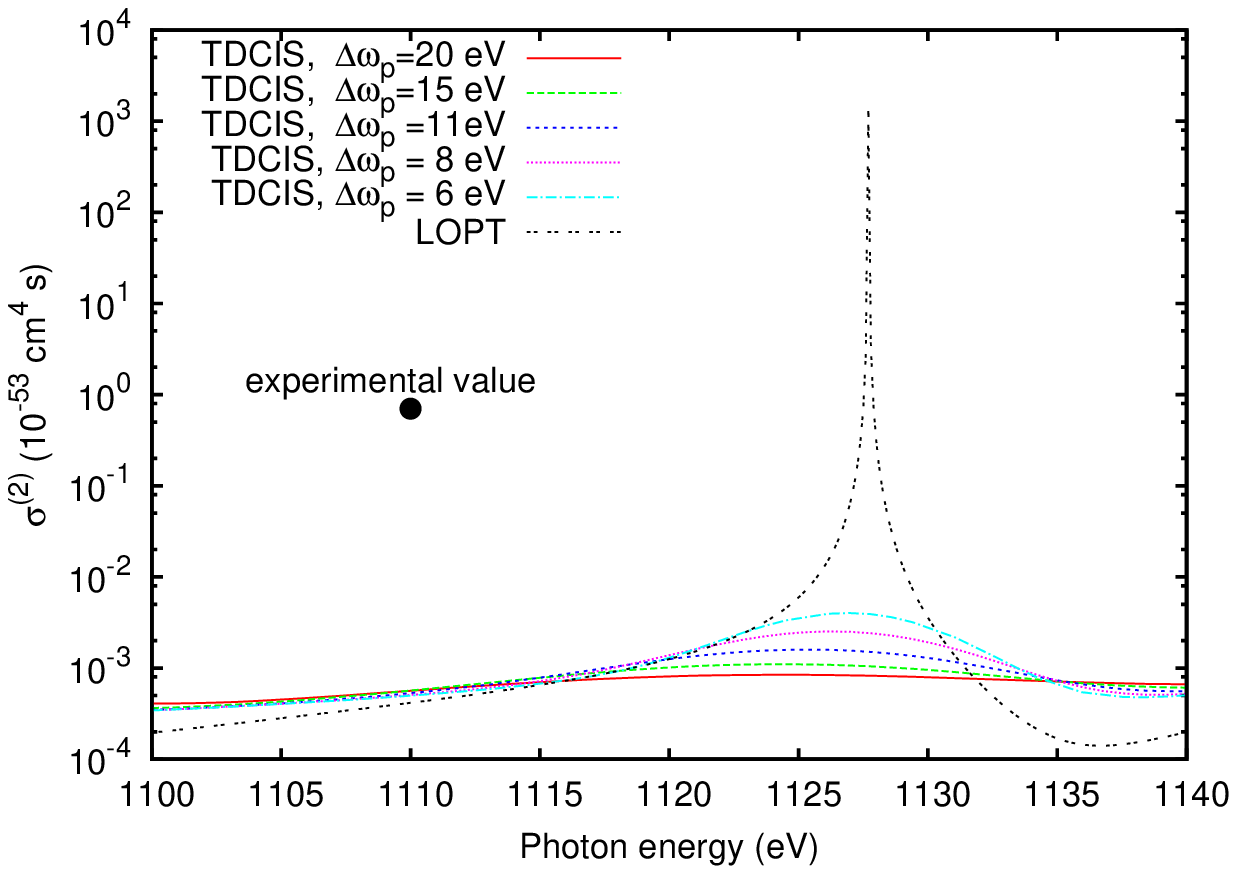} 
\caption{(Color online) Effective two-photon ionization cross section for Ne$^{8+}$. The TDCIS results are given by Eq.~(\ref{eq:cohcs}) for several different pulse bandwidths (FWHM). The LOPT result is obtained  using Eq.~(\ref{eq:lopt}). The point at 1110 eV corresponds to the experimental value of $7 \times 10^{-54}~\text{cm}^4 \text{s}$ reported in Ref. \cite{doumy-prl2011}.} 
\label{fig:ne8crosssec}
\end{figure}

\begin{figure}[h]

\centering
\begin{tabular}{cc}
\includegraphics{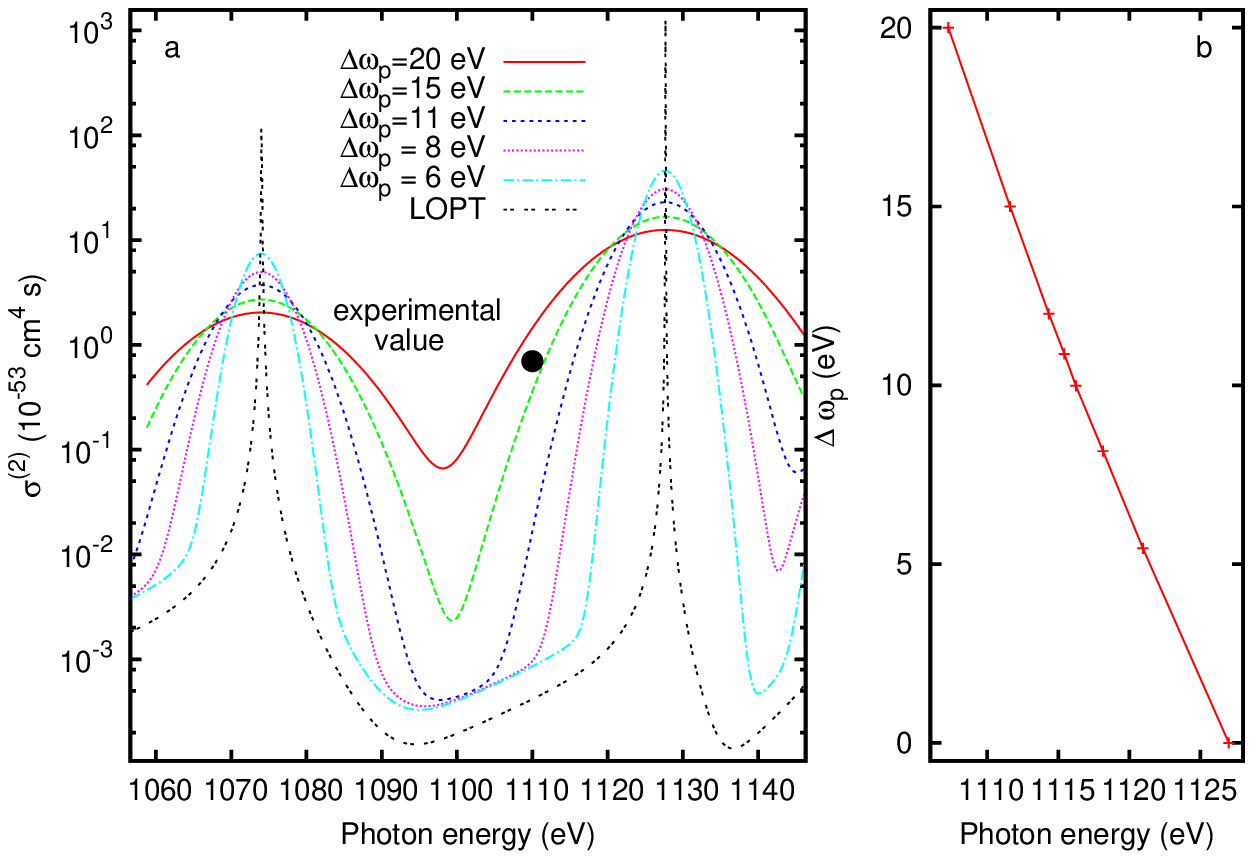} 
\end{tabular}

\caption{(Color online) (a) Two-photon ionization cross section for Ne$^{8+}$, given by Eq.~(\ref{eq:incohcs}). The perturbative result $\sigma^{(2)}_{\text{LOPT}}$ of Eq. (\ref{eq:lopt}) (dotted line) is taken as a reference signal for averaging over different bandwidths (FWHM) of the pulses. The point at 1110 eV corresponds to the experimental value of $7 \times 10^{-54}~\text{cm}^4 \text{s}$ reported in Ref. \cite{doumy-prl2011}. (b) Relation between the bandwidth $\Delta \omega_{\text{p}}$ and the mean photon energy $\omega_{\text{in}}$ for which the two-photon ionization cross section $\sigma^{(2)}_{\text{incoh}}$ is fixed at $7 \times 10^{-54}~\text{cm}^4 \text{s}$.}
\label{fig:lopt-conv}
\end{figure}

Doumy {\it et al.} \cite{doumy-prl2011} measured the mean photon energy with an uncertainty of several tenths of eV and the pulse spectral width was $10 \pm 1$ eV \footnote{G.~Doumy, (private communication).}. In Fig.~\ref{fig:ne8crosssec}, we show the two-photon ionization cross section, calculated using Eq.~(\ref{eq:cohcs}) for several pulse durations corresponding to the FWHM bandwidths of 20, 15, 11, 8 and 6 eV.  The peak electric field ${\cal E}_0$=0.03~a.u.\ was used. Also shown is the cross section $\sigma_{\text{LOPT}}^{(2)}(\omega)$ given by Eq.~(\ref{eq:lopt}). For the latter, we use the Hartree-Fock-Slater (HFS) model \cite{slater}, implemented within the \textsc{xatom} code \cite{xatom, son-xatom2011}. The HFS model positions the intermediate resonances at lower energies than those obtained in TDCIS, therefore we shifted the curve $\sigma_{\text{LOPT}}^{(2)}(\omega)$ such that the 1$s^2$--1$s4p$ resonance is at the right position of 1127.1 eV. Doumy {\it et al.} \cite{doumy-prl2011} noticed that in a similar perturbative calculation \cite{nov-hasp2001} the authors did not account for the $1s^2$--$1s4p$ resonance.  We have included this resonance in both TDCIS and LOPT calculations. However, as we see from Fig.~\ref{fig:ne8crosssec}, neither the inclusion of this resonance nor the finite bandwidth of the radiation pulse taken into account in TDCIS can explain the discrepancy of several orders of magnitude between the theoretical and experimental values.

Now, we use Eq.~(\ref{eq:incohcs}) to convolve the monochromatic two-photon ionization cross section obtained with Eq.~(\ref{eq:lopt}), with the spectral distribution function given by Eq.~(\ref{eq:distribution}), and show the results in Fig.~\ref{fig:lopt-conv}(a). One can see that within the bandwidth, off from the resonances, the cross section is substantially enhanced, because the main contribution to the convolution in Eq.~(\ref{eq:incohcs}) comes from the resonance peaks. Indeed, for a bandwidth of 11 eV the cross section at 1110 eV is $1.6 \times 10^{-55}~\text{cm}^4 \text{s}$, thus is enhanced by at least one and a half orders of magnitude with respect to the perturbative result ($4 \times 10^{-57}~\text{cm}^4 \text{s}$). In Fig.~\ref{fig:lopt-conv}(b), we show the relation between the pulse bandwidth $\Delta \omega_{\text{p}}$ and mean photon energy $\omega_{\text{in}}$, which is needed for the calculated two-photon ionization cross section to reach the experimentally found value of 7$\times 10^{-54}~\text{cm}^4 \text{s}$. For a bandwidth of $17$~eV, the calculated cross section increases up to this value at the photon energy of 1110 eV used in the experiment. Thus, our findings suggest that the main reasons for the enhanced two-photon ionization cross section of Ne$^{8+}$ at 1110 eV originate from the proximity of the 1$s^2$--1$s$4$p$ resonance, the chaoticity of the LCLS radiation, and the finite bandwidth of its pulses.

\begin{figure}[h]
\includegraphics[width=0.75\textwidth]{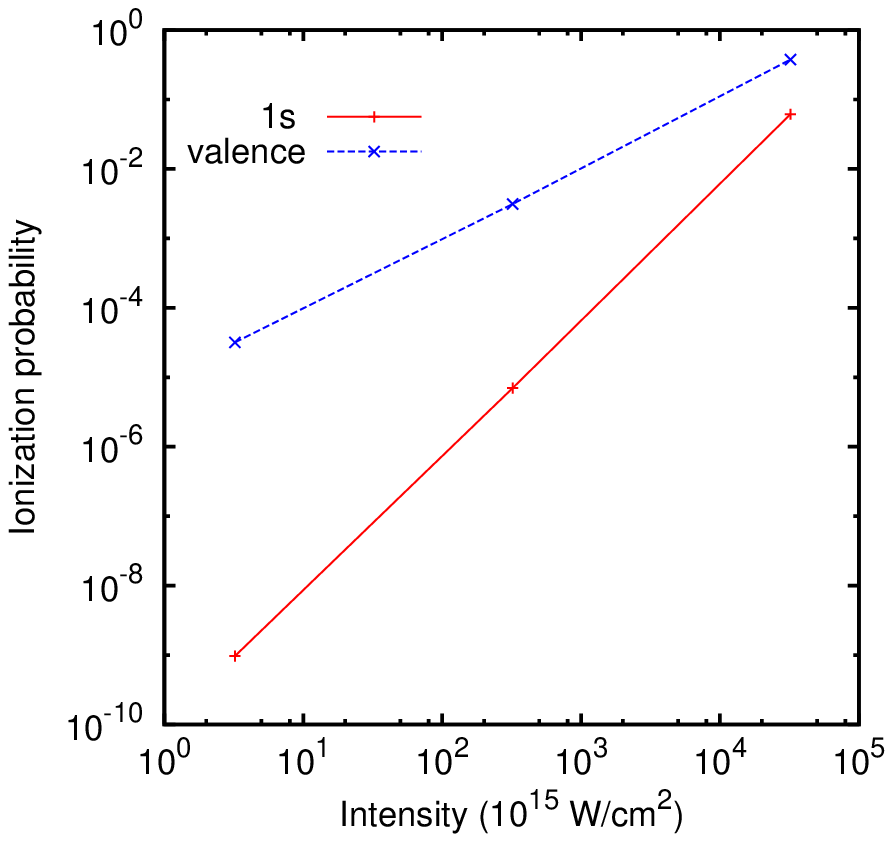} 
\caption{(Color online) Intensity dependence of the ionization probability of neutral neon, given by the IDM corrections $\delta \rho_{1s}$ for $1s$ electrons and $\delta \rho_{2s}+\sum_{m}\delta \rho_{2p_m}$ for valence electrons, at a photon energy of 800 eV (below the one-photon ionization threshold for the $K$-shell, but above the one-photon ionization threshold for the valence shells). A deterministic pulse of 6-eV bandwidth (FWHM) is used.}
\label{fig:neon_population}
\end{figure}

In connection with the study of two-photon ionization of core electrons, it is worth mentioning another recent experiment of Young {\it et al.} \cite{young-nature}, where direct multiphoton ionization of neon was completely shadowed by a sequence of one-photon ionization events. One of the measurements has been done at the photon energy of 800 eV, just below the $K$-edge, 870 eV, of neutral neon. In this case, one x-ray photon carries enough energy to ionize valence electrons, and therefore the valence-shell electrons are stripped in a sequence of one-photon absorption processes. Creation of a $1s$-shell vacancy is possible only through the absorption of two photons. No evidence for this process was detected. 

Using the TDCIS model, we study the possibility of creating a core hole in Ne via simultaneous absorption of two 800-eV photons. The converged result is obtained by using a maximum radius of 90 a.u. with 1000 grid points and $\zeta=0.461$. A CAP of strength  $\eta=0.002$ starts at 60 a.u. Accounting for angular momenta of the ionized electron up to $l_{\text{max}}=3$ is sufficient. In Fig.~\ref{fig:neon_population}, we show the ionization probabilities of valence and core electrons for neutral neon as a function of peak intensity. One can see that at the intensity of ~3 $\times$ 10$^{17}$ W/cm$^2$ the probability of ejecting a 1$s$ electron is more than 10$^2$ times smaller than that of ejection of a valence electron. With increasing intensity, the relative probability of 1$s$ ionization with respect to valence ionization grows. Nevertheless, this calculation shows that direct two-photon processes with ejection of an inner-shell electron never dominate the one-photon ionization of valence electrons, even for a pulse as short as $\tau_{\text{p}}=300$~as (corresponding to a bandwidth of 6 eV).  We confirm the observation of Young {\it et al.} \cite{young-nature} that multiphoton processes involving inner shell electrons are overshadowed by valence ionization as long as the valence electrons are not stripped away.

\section{Conclusion} \label{sec:conclusions}

In conclusion, we investigated the two-photon ionization cross section of Ne$^{8+}$ in the vicinity of the 1$s^2$--1$s4p$ resonance. We presented a strategy for calculating the two-photon ionization cross section within the TDCIS framework. However, the TDCIS model, which allows for a perfectly coherent radiation pulse, does not explain the enhanced two-photon ionization cross section, obtained by Doumy {\it et al.} \cite{doumy-prl2011} at 1110 eV, in spite of the inclusion of the 1$s^2$-1$s4p$ resonance which was missing in Ref.~\cite{nov-hasp2001}. The inclusion of the 1$s^2$-$1s4p$ resonance within the LOPT approach for monochromatic light does not explain the experimental result either. Chaoticity and short coherence time of the XFEL radiation, taken into accounted through the spectral distribution function in the cross-section expression obtained within LOPT, partially explain the observed enhancement. For the bandwidth of 10 $\pm~1$~eV, estimated in the experiment, we obtained an increase of the effective two-photon cross section by a factor of 40 with respect to the perturbative result for monochromatic radiation. To explain the experimentally observed value of $7 \times 10^{-54}~\text{cm}^4 \text{s}$ within this framework one would need a broader spectral bandwidth ($\sim17$~eV) or a mean photon energy tuned closer to the $1s^2$--$1s4p$ resonance. It is also worth noting that some indirect pathways that avoid production and two-photon ionization of ground-state Ne$^{8+}$ have not been included in the rate-equation model used in Ref.~\cite{doumy-prl2011}. This might have caused the experimental $\sigma^{(2)}$ to be overestimated. Nevertheless, we believe the $1s^2$--$1s4p$ resonance is the key to explain the enhanced two-photon ionization cross section of Ne$^{8+}$ at 1110~eV, but its influence depends strongly on the XFEL spectral density and uncertainties in its mean photon energy.

From the study of neutral neon performed within the TDCIS framework, we also infer that, when available, valence electron stripping due to one-photon ionization dominates over two-photon ionization of inner-shell electrons even at intensities far beyond current experimental possibilities.

\section{Acknowledgments}

We gratefully acknowledge helpful discussions with Joe Frisch and Linda Young.  
We thank Alexandro Saenz for e-mail communication regarding Ref. \cite{saenz}, and Gilles Doumy for providing important details on the experiment. 
A.S.\ is thankful to Gopal Dixit for his useful comments and Otfried Geffert for his help in software development.

\bibliography{references}

\end{document}